\RequirePackage[2020-02-02]{latexrelease}
\documentclass[prl,showpacs,preprintnumbers,amsmath,amssymb]{revtex4}
\usepackage[dvips]{graphics}
\usepackage{tikz}

\begin{document}

\title{A Matter-Wave Quantum Superposition of Inertial and Constant Acceleration Motions}

\author{Vlatko Vedral}
\affiliation{Clarendon Laboratory, University of Oxford, Parks Road, Oxford OX1 3PU, United Kingdom}

\date{\today}

\begin{abstract}
\noindent We present three different methods of calculating the non-relativistic dynamics of a quantum matter-wave evolving in a superposition of the inertial and accelerated motions. The relative phase between the two, which is classically unobservable as it is a gauge transformation, can be detected in a matter-wave interference experiment. The first method is the most straightforward and it represents the evolution as an exponential of the Hamiltonian. Based on the Heisenberg picture, the second method is insightful because it gives us extra insight into the independence of the wave-packet spreading of the magnitude of acceleration. Also, it demonstrates that the Heisenberg picture is perfectly suited to capturing all aspects of quantum interference.  The final method shows the consistency with the full relativistic treatment and we use it to make a point regarding the equivalence principle. 
\end{abstract}


\maketitle                           

\section{Introduction}
A particle subjected to constant force is one of the earliest exact calculations within proper non-relativistic quantum mechanics \cite{Kennard, Darwin}. It is a standard undergraduate problem \cite{Problems} whose solution is easily obtained. It also forms the basis of the W-K-B method for addressing a particle in a general potential \cite{Pauli}. We solve it in three different ways to address three different subtle points. The first point is the relationship between the motion in constant gravity (or acceleration) and free motion. Here we use the Schr\"odinger picture. The second point concerns the broadening of the wave-packet and the relative phase between the accelerated and free motion. This is most easily seen in the evolving position operator in the Heisenberg picture. The final point will be obtaining these results from the full relativistic treatment. Here we would like to emphasize how the equivalence principle is manifested in quantum physics. It is this point that we discuss more extensively in the final section.

\section{Solution By Exponentiation} 

Here is the simplest way of exponentiating the Hamitonian:
\begin{equation}
H= \frac{p^2}{2m} + mgx
\end{equation}
where $p$ and $x$ are operators and the rest are c-numbers having the usual meaning.
Let us introduce the following notation: $A = mgx$ and $B = p^2/2m$. The Zassenhaus expansion is (up to the third order as the higher orders vanish in our case)
\begin{equation}
\exp\{(A + B)\tau \}=   \exp\Big(A\tau \Big) \exp\Big(B\tau \Big)\exp\Bigg(-\frac{t^2}{2} [A,B]\tau \Bigg) \exp\Bigg(\frac{t^3}{6}([B,[A,B]]+[A,[A,B]])\tau \Bigg)
\end{equation}
Therefore
\begin{equation}
U=e^{-\frac{i}{\hbar} Ht} = \exp\Bigg(-\frac{i t}{\hbar}(\frac{p^2}{2m} + mgx) \Bigg)=   \exp\Bigg(-\frac{i t}{\hbar}mgx  \Bigg) \exp\Bigg(-\frac{i t}{\hbar}\frac{p^2}{2m}  \Bigg) \exp\Bigg(-i\frac{gt^2}{2} p\Bigg) \exp\Bigg(-\frac{i mg^2t^3}{6\hbar}\Bigg)  
\end{equation}
Applying this to a general initial state we obtain
\begin{equation}
\Psi (x,t) = e^{-\frac{i}{\hbar} Ht} \Psi (x,0) =  e^{-\frac{i mg^2t^3}{6\hbar}}  e^{-\frac{i t}{\hbar}mgx} e^{-\frac{i t}{\hbar}\frac{p^2}{2m}}e^{-i\frac{gt^2}{2} p}   \Psi (x,0) = e^{-\frac{i mg^2t^3}{6\hbar}}  e^{-\frac{i t}{\hbar}mgx} e^{-\frac{i t}{\hbar}\frac{p^2}{2m}} \Psi \Big(x + \frac{gt^2}{2},0\Big)
\end{equation}
This formula has a simple interpretation. The state of the particle falling freely in gravity is like that of a free particle whose position is shifted to $x+\frac{gt^2}{2}$ and an additional phase factor $e^{-\frac{i mg^2t^3}{6\hbar}}  e^{-\frac{i t}{\hbar}mgx}$. These phase factors are here global (and therefore seemingly undetectable), however, they can be observed if we use a reference state that is not accelerated and with which the accelerated state has interfered (which was proposed by Marletto and myself \cite{MAVE2} and has now been experimentally confirmed in \cite{Ron}). This leads us to another subtle point.

\section{Two-time Heisenberg Formalism}

One surprising aspect of the accelerated motion is that the acceleration does not affect the spreading of the wave-packet. The wave-packet spread is independent of $g$, i.e. it is the same as for the free particle. This follows from the above treatment, but can more conveniently be seen from the Heisenberg equations of motion. Namely, the position operator satisfies the following equation of motion
\begin{equation}
x(t) = x(0)+\frac{p}{m}t+\frac{gt^2}{2}
\end{equation}
where the momentum operator $p$ is time-independent ($p(t)=p(0)=p$). Taking the commutator with $x(0)$ of both sides of this equation leads to
\begin{equation}
[x(t),x(0)] = [x(0),x(0)]+\frac{1}{m}[p,x(0)] t+\frac{[g,x(0)]t^2}{2} = -\frac{i\hbar t}{m}
\end{equation}
The first commutator vanishes since it is the operator with itself and the third vanishes because $g$ is a c-number. This implies that 
\begin{equation}
\langle \Delta x(0)\Delta x(t)\rangle \geq \langle [x(t),x(t)]\rangle  = \frac{\hbar t}{m}
\end{equation}
from which we see that the spread at time $t$ is equal to $\frac{\hbar t}{m\Delta x(0)}$, exactly the same as for free motion. 

However, as we have seen above, there is a phase difference between the free and accelerated motion. To see how the phase can be obtained from the position operator, we can do the following. We will specify position operators at two different times $t_0$ and $t_1$. Then, the position operator at any other time $t$ is given by 
\begin{equation}
x(t) = \frac{x(t_0)t_1-x(t_1)t_0}{t_1-t_0} -\frac{gt_0t_1}{2} +\Bigg(\frac{x(t_1)-x(t_0)}{t_1-t_0} + \frac{g(t_0+t_1)}{2}\Bigg)t + \frac{gt^2}{2}
\end{equation}
We now compute the two-time action for this (quantum) trajectory
\begin{equation}
S_g[x(t)] = \int_{t_0}^{t_1} \Bigg(\frac{m\dot x}{2} - mgx\Bigg) dt= \frac{m}{2}\Bigg(\frac{(x(t_0)-x(t_1))^2}{t_1-t_0} - g (x(t_0)+x(t_1))(t_1-t_0) - \frac{1}{12}g^2 (t_1-t_0)^3\Bigg)
\end{equation}
To obtain the evolution we need to calculate the amplitude $\langle x, t|S|x_0, t_0=0\rangle$ and contrast this with the corresponding amplitude for the free particle shifted by $\frac{gt^2}{2}$. Here we have that 
\begin{equation}
S_{g=0} [x+\frac{gt^2}{2}] = \frac{m}{2t}\Bigg(x(0) - x(t) + gt^2/2\Bigg)^2
\end{equation}
The relative propagator is therefore
\begin{equation}
\Delta S_{g} = S_g - S_{g=0} = -mg x(t) t - \frac{mg^2t^3}{6}
\end{equation}
which is exactly the result obtained by direct computation earlier. 

We mention in passing that the treatment in this section completely invalidates the claims that interference cannot be understood in the Heisenberg picture and by local means (i.e., by operators that pertain to the particle itself). The operator formalism contains all possible amplitudes for all possible interference experiments between all eigenstates of $x(0)$ and $x(t)$. Schr\"odinger's picture is, therefore, a special case of the Heisenberg picture \cite{Frenkel}, and both of course ultimately contain the same information.  

\section{From Klein-Gordon in Curved Spacetime}

Since gravity enters only as a c-number, the full relativistic treatment using the Klein-Gordon equation in curved spacetime should reduce to the non-relativistic treatment in the limit of low velocities. We demonstrate that this is the case. The metric (which we can think of either as due to non-gravitationally induced acceleration or genuine gravity, is given by $ds^2 = (1-gx/c^2)dt^2 - dx^2$ and therefore the Klein-Gordon equation assumes the form
\begin{equation}
g_{\mu\nu} \nabla_\mu \nabla_\nu \Phi = \frac{m^2 c^2}{\hbar^2} \Phi
\end{equation}
where $g_{\mu\nu}$ follows from the form of $ds^2$. The non-relativistic equation is obtained by making the following substitution: 
\begin{equation}
\Phi (x,t) = e^{imc^2t/\hbar} \Psi (x,t)
\end{equation}
and then proceeding to exclude terms of the order $c^2$ and higher. This leads us directly to the Schr\"odinger equation for $\Psi $. 
\begin{equation}
-\frac{\hbar^2 \partial^2}{2m\partial x^2}\Psi (x,t) +mgx \Psi (x,t) = i\hbar \frac{\partial}{\partial t} \Psi (x,t) 
\end{equation}
Note that once we make a superposition of the free-falling matter wave with a stationary one, the roles of the two could be switched in our description and the result will be the same. The reason for this is the equivalence principle which states that gravity could always be gauged away by moving to the freely falling frame. In that frame, the wave in the falling branch is inertial, while, in the other branch, it is accelerating. The equivalence principle is manifested as a local gauge and, as such, we can only detect the relative phase between the branches. In other words, acceleration (or gravity) cannot be gauged away in both branches simultaneously
(This is true for any relative phase between two orthogonal states $|\psi_1\rangle + e^{i\phi}|\psi_2\rangle$; it can be written to appear to pertain to one state or the other by simply multiplying to total state by $e^{-i\phi}$ to obtain $e^{-i\phi}|\psi_1\rangle + e^{i\phi}|\psi_2\rangle$). 

The relativistic treatment reveals a different explanation of the relative phase, which is all about the ``curvature of time". The action can now be written as 
\begin{equation}
S = mc^2 (T-t) = mc^2 \Bigg(\int_0^t \sqrt{(1-gx/c^2) - (\frac{dx}{dt})^2}dt - t\Bigg)
\end{equation}
Expanding this to the first order gives us the non-relativistic result obtained before:
\begin{equation}
S \approx - mgxt - \frac{mg^2 t^3}{6}
\end{equation}
If $x$ and $p=mdx/dt$ are treated as operators, the resulting action would be an operator and we would have to sandwich it between the initial and final states to obtain the c-number amplitude. This is, in fact, the centerpiece of Schwinger's approach to quantum physics \cite{Schwinger}.  Needless to say, this relativistic approach leads to the correct phase difference to all orders in $g$. The spreading of the relativistic wave-packet can be studied in the same way as in the previous section and will deviate from the non-relativistic wave-packet for the simple reason that its speed can never exceed the speed of light. 

\section{Discussion}

We have presented three different ways of treating a single wave-packet of matter undergoing the superposed accelerated and inertial evolutions. Here we would like to discuss the full experiment which includes the creation of the superposed state. Suppose our matter wave is a Bose condensate whose spin degree of freedom can be manipulated. We can then perform a standard Stern-Gerlach experiment in which, when the spin is in the state $|0\rangle$ of the $Z$ component, the condensate is sent on a ballistic trajectory in Earth's gravitational field. At the same time, if the spin is in the state $|1\rangle$, the condensate is left stationary. The overall evolution is given by the unitary transformation
\begin{equation}
U = |0\rangle\langle 0| \otimes U_g + |0\rangle\langle 0| \otimes U_{g=0} 
\end{equation} 
where $U_g$ and $U_{g=0}$ are the propagators for the wave-packet in gravity and no gravity respectively. After a certain time, $t$, these two branches interfered by bringing them together back to the same spatial location. The interference fringes are obtained as the expected value of the spin $X$ component. 

We would now like to conclude by commenting on the Equivalence Principle in the quantum domain. In that context, we have argued that other principles originally formulated within a classical theory, such as the Einstein equivalence principle, should be extended to quantum systems in exactly the way outlined above, \cite{MAVE1, MAVE2}. Stated more generally, if we have a spatial superposition in which a massive particle exists at two locations, then this generates two different gravitational fields at a given distant point (assuming that the gravitational field is treated quantum-mechanically and in the first linear order of approximation. A test particle located at that distant point would then accelerate in both branches, towards the massive particle's respective locations. The state of the initial mass, the field and the test mass would then be:
\begin{equation}
\frac{1}{\sqrt{2}}\left (|r_1\rangle |g_1 \rangle |a_1 \rangle +|r_2\rangle |g_2\rangle |a_2 \rangle\right)
\end{equation}
The equivalence principle, which says that the gravitational field is indistinguishable (locally) from acceleration, applies in each of the superposed spatial branches. This is to be expected since each branch represents a classical gravitational scenario, where the position observable is sharp with the respective values. It is of course possible that this view of the equivalence principle will be experimentally invalidated (we do not have any experimental evidence in this domain to guide us), however, our point is that there is no prima facie reason to think that the equivalence principle is in conflict with quantum physics (any more than energy conservation is). 

Any consistent account of classical conservation laws and symmetries must be handled in quantum theory by incorporating all the relevant degrees of freedom so that the conservation and symmetry become exact, expressed at the level of q-numbers, or branch-by-branch.  This mandates to include in our models all the relevant entanglements due to the interaction between the system and the environment, including measurement apparatuses \cite{MAVE1, MAVE2}. In that sense, it is possible that when the full quantum gravitational treatment is done, superposing two large accelerations or two large masses will not be possible due to the resulting entanglement with the gravitational field. None of this, however, is relevant in our domain of low energies investigated in this paper.

\textit{Acknowledgments}: The author thanks Chiara Marletto for the extensive discussions related to the topics explored in this paper. VV's research is supported by the Gordon and Betty Moore and Templeton Foundations.

\end{document}